\documentclass[preprint2]{aastex631}

\defcitealias{2024Richardson}{\scshape R24}
\defcitealias{2012Klassen}{\scshape K12}

\begin{document}

\title{Introducing PxP: A Population Synthesis Framework for Predicting YSO Properties}

\correspondingauthor{J.~Peltonen}
\author[0000-0002-5937-9778]{J.~Peltonen}
\email{peltonen@ualberta.ca}
\affiliation{Department of Physics, University of Alberta, Edmonton, AB T6G 2E1, Canada}
\author[0000-0002-5204-2259]{E.~Rosolowsky}
\affiliation{Department of Physics, University of Alberta, Edmonton, AB T6G 2E1, Canada}

\author[0000-0001-6431-9633]{A.~Ginsburg}
\affiliation{Department of Astronomy, University of Florida, PO Box 112055, USA}

\author[0000-0002-4663-6827]{R.~Indebetouw}
\affiliation{National Radio Astronomy Observatory, 520 Edgemont Road Charlottesville, VA 22903, USA}
\affiliation{Department of Astronomy, University of Virginia, P.O. Box 3818, Charlottesville, VA 22903-0818, USA}

\author[0009-0001-8880-6951]{T.~Richardson}
\affiliation{Department of Astronomy, University of Florida, PO Box 112055, USA}

\author[0000-0002-0579-6613]{M. Jimena Rodr\'iguez}
\affiliation{Space Telescope Science Institute, 3700 San Martin Drive, Baltimore, MD 21218, USA}
\affiliation{Instituto de Astrofísica de La Plata, CONICET--UNLP, Paseo del Bosque S/N, B1900FWA La Plata, Argentina }



\begin{abstract}
The most direct method of measuring the star formation rate is with young stellar objects (YSOs), but this requires high-resolution observations and high-quality models. Using the latest YSO radiation transfer and stellar evolution models, we have developed a population synthesis code that generates model YSO populations that can be observed by JWST. We combine these model populations with principal component analysis (PCA) and maximum likelihood fitting to create a complete framework for predicting the age and mass of YSO populations. We dub this combination of Population synthesis and PCA, PxP, and show that it is effective at predicting mass and age with self-fitting tests. We apply PxP to the \textit{Spitzer} identified YSOs in N44 and find a mass of (1.1$\pm0.1$)$\times10^4$~M$_\odot$ and an age of $0.74^{+0.06}_{-0.03}$~Myr, consistent with previous work. Next, we identify 112 YSO candidates in the archival JWST observations of NGC~604. Applying PxP to this newly identified population we find a mass of $(2.2\pm0.2)\times10^4$~$M_\odot$ and an age of $0.62\pm0.01$~Myr. This first look at this framework demonstrates its effectiveness with a specific set of models and leaves clear opportunities for future exploration. PxP allows us to directly determine the recent ($<$3~Myr) star formation history, giving an unprecedented look at the effect of the large-scale environment on individual star formation.
\end{abstract}



\section{Introduction} \label{sec:intro}

To understand the evolution of galactic systems we must understand if and how star formation changes with galactic environment. One of the most natural quantities to quantify star formation is the star formation rate, or the mass of stars produced within some time. A related quantity is the star formation efficiency of the molecular gas where the stars are forming. The star formation efficiency links the star formation rate to the mass of the molecular gas. There are many ways to indirectly measure the star formation rate (e.g., H$\alpha$), that infer the star formation rate from the intense radiation output of short-lived, high mass stars, which all come with various systematic uncertainties \citep{2012Kennicutt}. However, these stars are necessarily separated, both temporally and spatially, from the stars that are forming in molecular clouds. Despite these uncertainties, star formation observations have converged to a near universal star formation efficiency per free-fall time with some variation with molecular gas properties and galactic environment \citep{2025Leroy}.

The best way to determine the star formation rate would be to directly measure the mass and age of forming stars known as young stellar objects (YSOs). To actually observe YSOs, one requires superb resolution, and to determine their mass and age, equally superb models. These observations are typically made with infrared observatories that can peer into the high extinction clouds where the YSOs are forming. This has left YSO studies to mostly be conducted in nearby Milky Way clouds \citep[e.g.,][]{2010Lada,2015Dunham}. Sampling a wider range of Galactic environments is difficult due to extinction and distance ambiguities \citep[e.g.,][]{2009Roman, 2018Motte}. Studies of the Magellanic clouds using \textit{Spitzer} \citep[e.g.,]{2008Whitney,2013Sweilo} were the first studies that could study massive YSOs and relate those forming stars to their host molecular clouds \citep{2017Ochsendorf,2022Finn}

The James Webb Space Telescope (JWST) has allowed for the observation of YSOs far beyond the Milky Way. JWST is able to identify massive YSOs as far as $\sim$1~Mpc. For example, \citet{2024Lenkic} were able to identify YSOs in the Spitzer I region of NGC~6822 at a distance of $\sim$500~kpc. \citet{2024Peltonen} were able to identify $\sim 10^3$ massive YSO candidates in the south-west region of M33 at 859~kpc \citep{2017distanceM33}. In addition to massive YSOs at greater distances, JWST also allows for less massive YSOs to be observed in the Magellanic clouds \citep{2024Habel}. JWST has opened the door to observing YSOs in a much broader range of environments than ever before.

To tie these JWST observations of YSOs to star formation rates, we require models that predict the age and mass. The radiative transfer models of \citet{2006Robitaille} were widely used in the \textit{Spitzer} era, as they would simply take observed YSO spectral energy distributions (SEDs) and estimate the age and mass. However, these models are inflexible and cannot adjust accretion history, unlike the updated \citet{rob17} and \citet[][\citetalias{2024Richardson}]{2024Richardson}. These models all attempt to determine the properties of individual YSOs from observed SEDs. Thus, if only limited SED is available, the predicted parameters can be uncertain. One method that has been utilized to overcome this uncertainty is fitting to multiple sources instead of a single source. This approach is known as population synthesis and has been used to great effect in main-sequence populations with codes like \texttt{MATCH} \citep{2002Dolphin}.  In the context of YSO studies, \citep{2003Muench} proposed a model for inferring initial mass function (IMF) properties from the observed luminosity distribution.

Our goal is to infer the short timescale ($<3~\mathrm{Myr}$) star formation history from the distribution of fluxes of unresolved sources in multiband JWST imaging data. This focuses on modeling the sources that are infrared-bright and red. These sources could be individual massive YSOs, young clusters, or possibly contaminants like AGB stars and background galaxies.  Our approach uses an ensemble of early-time star formation histories to create a set of mock catalogs of this region. We then determine which star formation histories are most consistent with observations. 

In section 2, we describe our population synthesis code. Section 3 describes how we use Principal Component Analysis (PCA) to fit our model population to real populations. This method, which we dub PxP, is tested in Section 4 using self-fitting and a known YSO population. Section 5 focuses on JWST observations of NGC~604 where we first identify YSO candidates and then determine their properties using PxP. Finally, Section 6 presents a summary of our results.

\section{Population Synthesis Code}\label{sec:popsyn}
\begin{figure*}
\includegraphics[width=\textwidth]{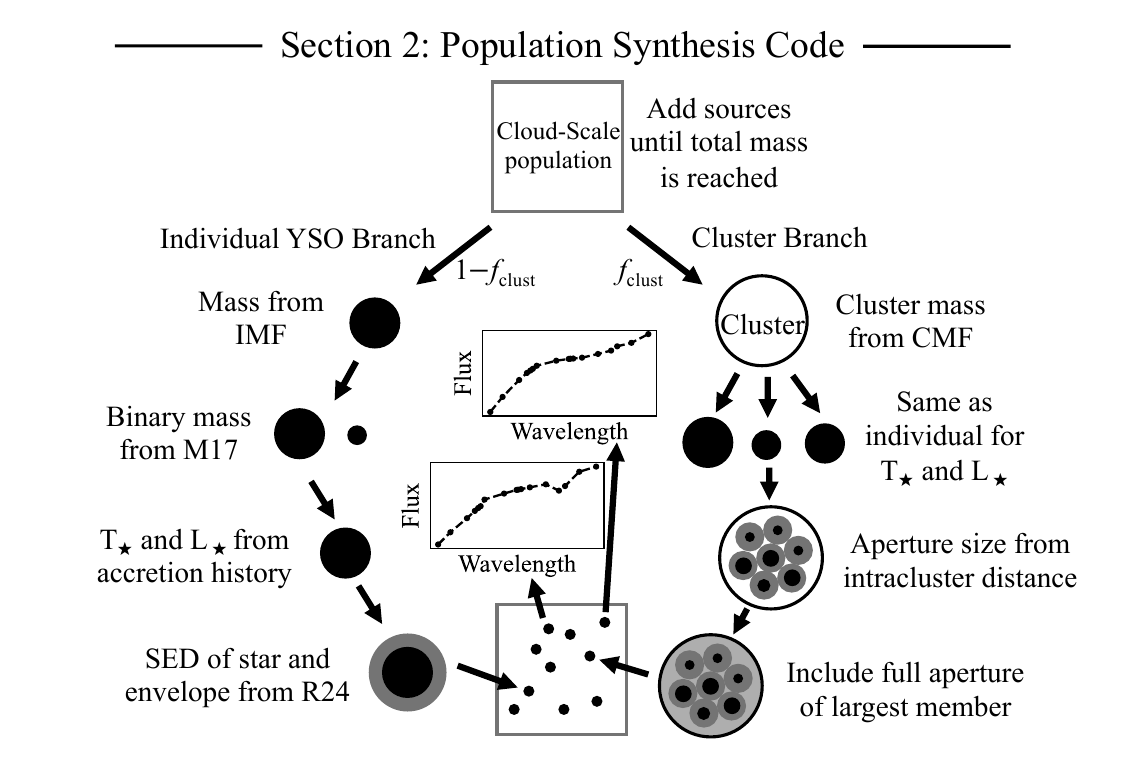}
\caption{Illustration of the population synthesis hierarchy proposed. The starting point is the age and mass of the cloud-scale population. Both individual YSOs and unresolved clusters are then added until the total mass is reached. Individual YSO SEDs are determined by following the left path and unresolved cluster SEDs are determined by following the right path.}\label{fig:hierarchy}
\end{figure*}

We build our model of an ensemble of YSO sources using a set of assumed properties of the forming stellar population and then use grids of radiative transfer models for YSOs to convert these into a set of source SEDs. Figure \ref{fig:hierarchy} schematically illustrates the relationships between the components of our model. To be clear about our terminology, \textit{Populations} refer to cloud-scale ($\sim100$~pc) collections of observable pointlike \textit{sources}. These sources can be either \textit{individual YSOs}, which will end up as single stellar systems, or \textit{unresolved clusters}, which are small ($\sim$1~pc) collections of YSOs that appear like point sources at the distances to the objects (e.g., MIRI at $\sim1$~Mpc). In this work, we adopt specific, simple models to describe the YSO population, but the general fitting framework we develop (Section \ref{sec:fitting}) can readily incorporate a different set of assumptions.

Our approach requires a model of: (1) the mass distribution of stars formed in the region, including the binary mass fraction distribution, the fraction of those star systems found in clusters, and the cluster mass distribution function; (2) the star formation history and the accretion history of the forming stars, including how that accretion history affects the temperature and luminosity of the protostars; and (3) the properties of the dust envelopes around these forming stars. We describe our assumed model in more detail below. Ultimately, we can create realizations of young stellar populations that depend primarily on the total mass of stars formed, the star formation history, and the foreground extinction distribution toward that population.

\subsection{Stellar Mass Distributions}

To generate one of these cloud-scale populations, individual YSOs and clusters are assigned to the population with masses drawn from the Initial mass function (IMF) and cluster mass function (CMF) until the total population mass is reached. We use the IMF from \citet{2001Kroupa} that gives the probability distribution of individual star masses. Similarly, the CMF gives the probability distribution of cluster masses. The shape of the CMF is defined as
\begin{equation}
        \frac{dN_c}{dM_c} \propto M_c^{-2} \exp\left(-\frac{M_c}{M_0}\right),
\end{equation}
which is adopted from \citet{slug4}. The cluster fraction, $f_\mathrm{clust}$, represents the fraction of stellar mass formed in unresolved clusters. The cluster fraction is an adjustable parameter that depends on the linear resolution of the observations. The populations considered below have relatively coarse resolution ($\sim 1$~pc), resulting in the small clusters being unresolved. Therefore, the cluster fraction used here is high, reflecting that most stars form in clusters and those clusters will be unresolved \citep[$f_\mathrm{clust,0} = 0.9$;][]{lada_araa}. In applying these methods to better resolved stellar populations (e.g., for the Solar Neighborhood) the parameter should be lower ($f_\mathrm{clust}\sim 0$). The clusters are made up of single YSOs drawn from the IMF until the cluster mass is reached. Using the binary population distributions from \citet{2017Moe}, some individual stars were given a secondary companion. \citet{2017Moe} present a meta-analysis of a wide range of binary observations that recommends parameters for the properties of the different stars in stellar systems (e.g., the fraction of systems that are binaries and multiples, the number of stars in those systems, their mass ratios, and orbital periods). The recommended distributions indicate that more massive stars are more likely to have a companion, for example, 40\% of 1~M$_\odot$ stars have a companion while 85\% of 10~$_\odot$ stars have a companion. \citet{2017Moe} find that the observed masses of the two stars in a binary are typically similar to the masses one would get from randomly drawing a pair from the IMF, with some minor discrepancies from a wide range of effects. Therefore, the lower mass binary companion typically contributes an insignificant amount of flux to a source. However, we want to accurately predict the total mass of these populations, thus, we include the secondary to contribute additional mass but relatively little flux.

\subsection{Star Formation and Accretion Histories}

Our model requires, as an input, the amount of stellar mass that begins forming as a function of time in the analyzed region, $M_\star(t)$. Following the language of population synthesis codes, we refer to this as the ``star formation history'' of the cloud, but note that it has a few subtleties related to the current model. Because we model the formation of stars and clusters, the time variable in the model represents the time at which a given mass of stars begins forming out of molecular gas, which means our sources will stop accreting at time $t>0$ and that the accretion time will depend on the accretion model adopted.  Moreover, the mass of the forming stars is changing over time so $M_\star$ in this star formation history refers to the final mass of the stellar population once it arrives on the main sequence.

For this initial exploration, we consider two simple cases of star formation history. The first case is the burst model where all members of a population begin accreting at the same moment in time. The other case is the continuous star formation history, where the ages of the individual sources is evenly distributed from the beginning of accretion to the current age of the population. Thus, the age of the continuous population represents the age of the oldest source and contains many younger sources as well. We will define the average star formation rate, $\dot{M}_\star$, for both of these star formation histories to be $M_\star$ divided by the age of the population.

Realistic star formation histories are thought to be complex and may depend on the spatial distribution and masses of the YSOs \citep{2018Getman,2023Grudic}. These two simple model star formation histories should effectively represent two extremes, bounding the behavior of the true star formation history.

With sufficient numbers of objects, these simplified star formation histories can be relaxed.  The model could be used to explore the duration of star formation in different regions and has the flexibility to statistically test whether high mass stars form in clouds after or before lower mass stars.  In the absence of these more sophisticated star formation histories, the simple model will introduce systematic uncertainty into the age estimations. We will explore more sophisticated star formation histories in future work.

To establish the accretion history of individual stellar systems, we use a modified version of the \citet{2012Klassen} stellar evolution code, where all primary stars were assigned a temperature and total luminosity depending on the stellar mass, age, and an adopted accretion prescription. The stellar evolution code gives both the intrinsic thermal luminosity of the central source and the total luminosity, which includes accretion luminosity. Following \citet{2025Richardson} we utilize the total luminosity to fully capture the state of YSO. \citetalias{2012Klassen} implemented singular isothermal sphere accretion \citep{1977Shu}, and this approach was modified by \citet{2025Richardson} to include competitive \citep{1997Bonnell,2001Bonnell} and turbulent-core \citep{2002McKee,2003McKee} accretion histories using the parameterization of \citet{2011Offner}. This parameterization defines the mass accretion rate that depends on several factors. For isothermal sphere accretion a constant accretion rate is given based on the final mass of the star. However, for turbulent-core and competitive accretion, accretion scales up with the current mass of the (proto)star.  

As shown in \citet{2025Richardson}, turbulent-core and competitive accretion have comparable formation timescales (deviations within $\sim$1~Myr to complete accretion), whereas isothermal sphere accretion leaves massive stars accreting long after lower mass stars. We generated populations using all of the accretion histories, but our initial testing showed that the \citet{2025Richardson} modified turbulent-core accretion was the most effective at replicating observed sources.  We thus focus on populations using turbulent-core accretion for the remainder of this work. Richardson (in prep.) will delve into the differences between accretion histories in greater detail. Another important caveat is that the current implementation prescribes smooth variations in accretion, which differs from observed stochastic accretion \citep[e.g.,][]{2023Fischer}. Our framework is capable of handling other varieties of accretion history, such as stochastic, but we do not explore those here.

\subsection{Model SEDs}\label{sec:modelsed}

With populations of sources that have temperatures and luminosities, we determined the observable fluxes using the simulated JWST filters included in the \citetalias{2024Richardson} radiative transfer models. Specifically, we chose the ``\texttt{spubsmi}'', ``\texttt{s-pbsmi}'', and ``\texttt{sp--smi}'' geometries originally described in \citet{rob17}, which contains the simulated flux of 60000 models, each viewed from nine different angles. All three of these geometries include a central source and ambient ISM. \texttt{sp--smi} includes a disk, \texttt{s-pbsmi} includes an envelope, and spubsmi includes both a disk and envelope. These three geometries should effectively cover a wide range of YSO evolutionary states. However, the \citet{2024Richardson} radiative transfer models have no prescribed evolution or timescale, and we must rely on matching to an accretion history to determine the timescale of a \citet{2024Richardson} model.

The fluxes of the \citetalias{2024Richardson} models are connected to the temperature and luminosity of the sources in the cloud-scale population differently for isolated YSOs and unresolved clusters. For each isolated YSO, a \citetalias{2024Richardson} model was randomly selected from the 10 nearest neighbors in temperature-luminosity quantile space (quantile transform fit to the temperature and luminosity of all \citetalias{2024Richardson} models). A random inclination is chosen, then the flux of that star was taken to be the flux in the largest aperture ($10^6$~AU) in each filter, given by the chosen \citetalias{2024Richardson} model. 

For a given mass, the majority of the turbulent core accretion tracks occupy a unique part of the temperature-luminosity quantile space. However, there are small portions in temperature-luminosity space where accretion tracks overlap with other tracks of different masses in a different evolutionary state. Therefore, YSOs will occasionally be matched to an \citetalias{2024Richardson} model that more closely resembles a YSO of different mass and age then the one assigned. The effect these overlaps have on the overall population should be minor since the populations contain many members, and the effect will be further minimized by generating many populations with the same age and total mass.

The \citetalias{2024Richardson} models assume completely isolated sources and give simulated fluxes within different size apertures, the largest of which being $\sim5$~pc, larger than assumed size of our unresolved clusters. To determine the \citetalias{2024Richardson} for each cluster member, we used the same approach of selecting one of the ten nearest neighbors in temperature-luminosity space. However, including the flux of the largest apertures for all of the members would lead to overcounting flux contributions from overlapping envelopes. Therefore, we used the number of cluster members to determine the average member separation, assuming the cluster diameter is the size of the smallest MIRI resolution element (0\farcs207). Then, the flux from the aperture (in arcseconds) that matches that separation for each member was added to the total cluster flux. To estimate the flux contribution of the intra-cluster medium, we also included the largest aperture flux contribution for the most massive member of the cluster. The final cluster flux in each filter is found by adding the flux contributions of all of the members and the intra-cluster medium. Therefore, each source added to a cloud-scale population will add one SED regardless if it is an individual YSO or an unresolved cluster. We do not include any additional envelope clearing prescription for the intracluster gas. However, $\sim$85\% of the flux comes from the largest member for the majority of these clusters. This effect is wavelength dependent, with the largest member contributing at least 85\% of the flux for 68\% of the clusters at long wavelengths (21~$\mu$m) and 50\% of the clusters at short wavelengths (2~$\mu$m). This means that including the low-mass star contributions to these unresolved clusters affects the total mass while having only a modest effect on the total light. 

The \citetalias{2024Richardson} models contain local extinction effects from the envelope and ambient ISM; however, we also include the possibility of foreground extinction from the foreground Milky Way, the extragalactic environment, or the host molecular cloud. In principle, we can draw this extinction from a distribution for each population. In this initial implementation, we adopt a simple screen along the line of sight with a constant extinction across the entire population. We use the extinction parameterization of \citet{2023Gordon} based on average Milky Way measurements to convert a single value of visual extinction ($A_V$) to the extinction appropriate for each filter. This single $A_V$ value is included as a free parameter in generating a population.

\begin{figure*}
\includegraphics[width=\textwidth]{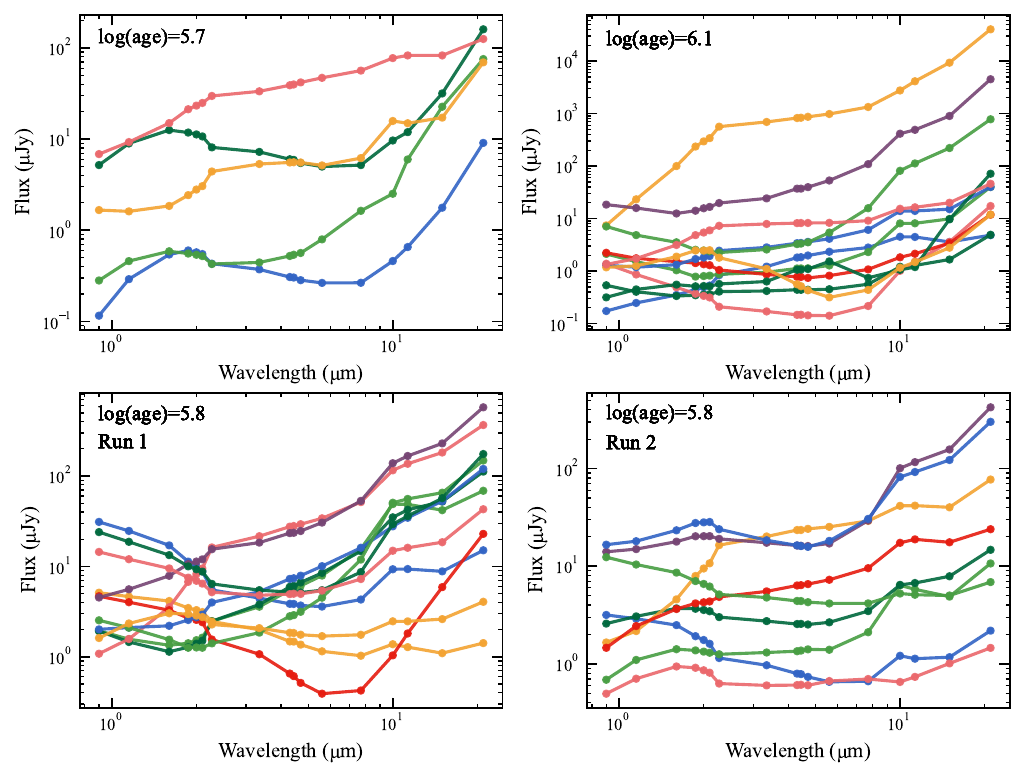}
\caption{Example output population SEDs from the synthesis code, where each coloured line represent a single source (isolated YSO or unresolved cluster) in the population. All populations are $10^3$~M$_\odot$, and the points show the synthetic JWST photometry with colours and connecting lines only to guide the eye. The top two panels show runs with two different ages, while the bottom two panels show two populations with the same properties. The notable differences between populations with the same properties illustrate the need for dimensionality reduction and a large grid of generated models to find the most likely properties of a real population. \label{fig:example_sed}}
\end{figure*}
\subsection{Example Results}

Under these assumptions, we can generate model populations with only three free parameters: total mass ($M_\star$), age ($t_0$, defined as time since the start of protostar formation), and foreground extinction ($A_V$). These cloud-scale populations will contain many source SEDs, where each SED corresponds to either an individual YSO or unresolved cluster. To compare these model populations to real populations, we must know which sources will be too dim to detect, which depends on the distance to the target and sensitivity of the observations. 

Here, we focus on M33 observations with JWST, which are comparable in terms of sensitivity, physical resolution, and wavelength coverage to observations of the LMC with \textit{Spitzer}. Therefore, all sources that are too dim to observe by JWST at a distance of $\sim$1~Mpc were marked as undetectable, and their SEDs are not included in the aggregate SED analysis (the mass of these undetectable sources will still contribute to the total mass of the population). Specifically, SEDs were excluded for sources lower than the minimum fluxes in the \cite{2024Peltonen} M33 YSO Candidate catalog (10~$\mu$Jy at 21~$\mu$m and 1~$\mu$Jy at 1.6~$\mu$m). Currently, our framework implements 17 commonly used JWST filters (F090W, F115W, F150W, F187N, F200W, F210M, F277W, F335M, F430M, F444W, F470N, F560W, F770W, F1000W, F1130W, F1500W, and F2100W), but can be easily modified to include additional or alternative filters. It should be noted that some of these filters may not be accurate representations of observations, specifically F335M, F770W, and F1130W, which are often dominated by polycyclic aromatic hydrocarbons that are not well simulated in the \citetalias{2024Richardson} models. Four example population SEDs from this code are shown in Figure \ref{fig:example_sed}, which illustrates the diversity of populations with the same age-mass-extinction parameterization but different random realizations and also compares these distributions between different sets of properties. From only these four example populations, it can be seen that even populations generated with the same parameters can vary greatly. However, many of the sources have common features and overall SED shapes.

\begin{figure}
\includegraphics[width=\columnwidth]{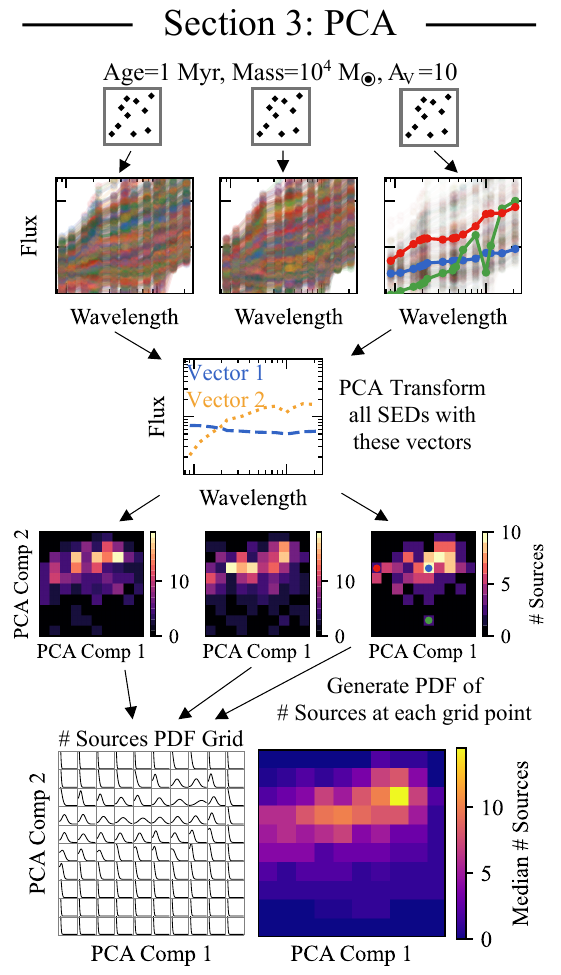}
\caption{Illustration of how we go from many iterations of a cloud-scale population with the same parameters into a single grid of PDFs. The right population has three of its SEDs highlighted (red, blue, and green lines), and where those SEDs are translated onto the PCA grid is marked with matching coloured points. This process is then repeated for each age, mass and extinction. \label{fig:PCA_info}}
\end{figure}

\section{Fitting}
\label{sec:fitting}

With this large set of populations with source fluxes, we required dimensionality reduction to fit age, population mass, and extinction. A common approach in population synthesis is to match the density of sources in optical color-magnitude diagrams \citep[e.g.,][]{2002Dolphin}. However, there is no single color-magnitude or color-color diagram in the infrared that can represent the complexity of the SEDs \citep{2017Jones}. Nonetheless, YSO colors span a small subspace within the different possible color-color diagrams.  To discriminate between different YSO populations, we perform dimensionality reduction through PCA, which reduces each SED into a weighted combination of principal component vectors \citep[see e.g.,][ for a complete discussion]{1987Murtagh}. 

We began by generating model populations with 11 masses logarithmically spaced from 10$^3$-10$^5$~M$_\odot$ and 50 ages linearly spaced from 0.3-3~Myr. Younger and less massive populations were not included since they lack significant numbers of sources above our flux threshold. For each of these masses and ages, we generated 1000 populations for a total of 550000 populations. We then applied a constant extinction screen to each of these SEDs with $A_V$ from 0-20 magnitudes for a total of 11,550,000 unique sets of SEDs. 

We determined the principal component basis using all of our source fluxes (the SED of all of the individual YSOs and unresolved clusters). Then, for each cloud-scale population, we determined the PCA weights of each SED in the population in the first two components since these first two components contained $\approx97\%$ of the variance in the full sample. As can be seen in Figure \ref{fig:PCA_info}, the first component vector (blue dashed line) is relatively flat and corresponds to the luminosity of the source SED. The second component vector (yellow dotted line) appears to represent the rising slope and depth of the $\sim10$~$\mu$m silicate feature. Using these principal components we reduce the full SED analysis into a two dimensional space. 

Figure \ref{fig:PCA_info} shows three realizations of a population generated using the same age, mass, and extinction. Each SED contained in a population is transformed using the PCA vectors, so two PCA weights can represent an SED. We then created 10$\times$10 two-dimensional (2D) histograms for each cloud-scale population. We adopted 10 bins to have sufficient resolution while avoiding small-number statistics. The three highlighted SEDs in Figure \ref{fig:PCA_info} show the physical interpretation of these 2D histograms. Moving from left (red source) to right (blue source) corresponds to a decrease in brightness. Moving from top (blue source) to bottom (green source) corresponds to a steeper rise to the red end of the SED with deeper silicate absorption. 

Given the stochastic nature of these models there is variation in the number of sources mapped to each grid point in the 2D histogram, even when using the same input parameters. Therefore, for each grid point, we fit a folded normal distribution with a probability density function (PDF) defined as 
\begin{equation}
    p(n) \propto \sqrt{\frac{1}{2\pi\sigma^2}}\left(e^{-\frac{(n-\mu)^2}{2\sigma^2}}+e^{-\frac{(n+\mu)^2}{2\sigma^2}}\right)
\end{equation}
where $\mu$ is the mean, $\sigma^2$ is the variance, and $n$ is the number of sources, restricted to $n\ge 0$.  We adopt the folded normal functional form based on a better empirical match to the population distributions than simpler forms like Poisson or normal distributions. 

Figure \ref{fig:PCA_info} shows an example distribution with 1000 populations with age=1~Myr, mass=$10^4$~M$_\odot$, and $A_V=10$, each with a unique number of sources at each grid point. The top left grid point has zero sources in the majority of the 1000 populations, which is well fit by a folded normal distribution centered at zero with a small variance. This process is repeated for each grid point, giving a grid of PDFs, where the PDF represents the probability of having a number of sources in a population with those PCA components. The visualization in \ref{fig:PCA_info} projects this grid of PDFs into a 2D histogram by showing the median number of sources in each bin. However, the final product for each age-mass-extinction combination is a grid of principal component weights, where each point in the space is a PDF of the expected number of sources with those PCA weights for a population. 

\begin{figure}
\includegraphics[width=\columnwidth]{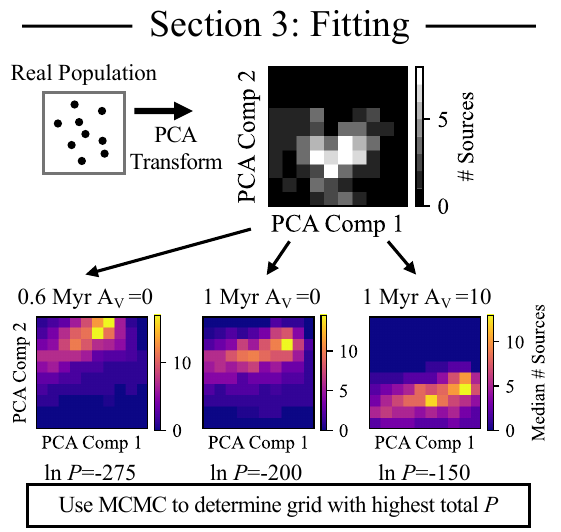}
\caption{Illustration of how we fit a real population to the best PDF grid. \label{fig:Fittinginfo}}
\end{figure}

With these grids of PDFs, as shown in Figure \ref{fig:Fittinginfo}, we can determine the matching properties of a new population of sources. We do this in a Bayesian approach where we infer $p(M_\star, t_0, A_V|\mathrm{PC}_1, \mathrm{PC}_2)$ where the PC$_i$ are the components generated from the observed data. We adopt flat priors over the range of parameters sampled in our models (see above). The probability that a real population is represented by grid of PDFs (visually represented by median 2D histograms in Figure \ref{fig:Fittinginfo}) can be found by summing the log-probabilities at each grid point using the actual number of sources and the model PDFs. We then estimate the posterior distribution of YSO population properties by sampling the total log-probability of the model parameters. This sampling is done using the Markov chain Monte Carlo package \textsc{emcee} \citep{2013emcee}. 

We will refer to the entirety of this approach as PxP since it combines \textbf{P}opulation synthesis with \textbf{P}CA to determine YSO population properties.

\section{Testing PxP}
To quantify the accuracy of PxP, we need to test on both model and real YSO populations. To quantify PxP's ability to successfully recover YSO population properties, we first show a series of self-retrieval tests. Next, to contrast PxP with previous attempts to measure observed YSO population properties, we apply PxP to N44, a well-studied star-forming region in the LMC.

\subsection{Self-Fit Testing}
First, we test PxP's ability to retrieve the properties of model populations. Using the same population synthesis methodology described in Section \ref{sec:popsyn}, we created new populations with a range of properties. We made the same flux cuts only including sources that would be detectable in the \citet{2024Peltonen} YSOC catalog. We then fit these new populations with PxP and compared the estimated properties to those that were used to generate the model population. 

\begin{figure}
\includegraphics[width=\columnwidth]{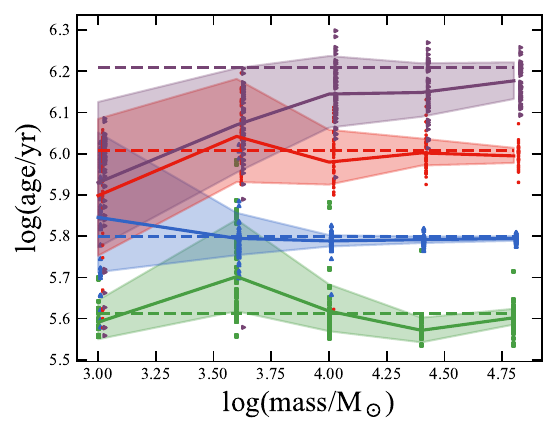}
\caption{The results of the self-fitting tests to determine the accuracy of PxP's age predictions. The newly generated populations were fit by PxP for four ages (green, blue, red, and purple) and five population masses. The points show the median age produced by PxP for each population with a small offset to input mass for clarity. The solid line shows the average median produced by PxP and the dotted line shows the input age used to generate the population. The shaded region shows the average interquartile range produced by PxP. The results of the self-fitting tests show that the age estimations of PxP are reasonable for populations above $\sim10^4$~M$_\odot$. \label{fig:agevm}}
\end{figure}

\begin{figure}
\includegraphics[width=\columnwidth]{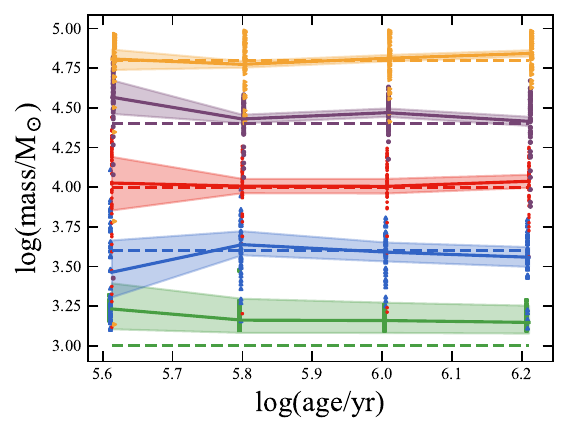}
\caption{The same as Figure \ref{fig:agevm} but showing the accuracy of PxP's mass predictions. The mass estimates of PxP are reasonable regardless of input age or mass. \label{fig:mvage}}
\end{figure}

As shown in Figure~\ref{fig:agevm} and \ref{fig:mvage}, PxP was able to recover the input parameters with relatively high precision if there are enough sources. However, lower mass populations with fewer sources do not give an accurate age prediction. The age predictions of PxP should not be used unless there are at least 15-40~sources, which translates to 4000~M$_\odot$ population at 1~Mpc. The interquartile range (IQR) of the predicted ages is $\approx0.2$~dex at 4000~M$_\odot$ and $<0.1$~dex for populations $\geq10000$~M$_\odot$. Figure~\ref{fig:agevm} also shows that older ages are less accurately predicted and populations older than $\sim2$~Myr are indistinguishable from one another for $M_\star < 10^4~M_\odot$. The total mass of the population is predicted more consistently than age. As can be seen in Figure \ref{fig:mvage}, for very young ages ($\leq$0.4~Myr), the mass predictions become less accurate because the number of sources is significantly affected by stochasticity, as fewer sources have reached high luminosities. For older ages ($>0.4$~Myr), the IQR is consistent at $0.2$~dex for low mass to $0.05$~dex for high masses. We note that populations of $1000$~M$_\odot$ are predicted to have a systematically higher mass, which is due to $1000$~M$_\odot$ being the minimum allowed value in our flat mass prior.

\subsection{N44}
We now apply PxP to N44, a star-forming region in the LMC. \citet{2012Carlson} created the most complete YSO catalog utilizing \textit{Spitzer} SEDs, spectroscopy, and ancillary optical observations. The 139 identified YSOs in N44 were fit with \citet{2006Robitaille} models to determine their individual masses. Next, \citet{2012Carlson} determined the total YSO population mass by fitting the high-mass end of a \citet{2001Kroupa} IMF to the YSO masses. This method yielded an estimated total mass of the N44 YSOs to be $\sim$3537~M$_\odot$. \citet{2009Chen} fit each YSO that they identified with \textit{Spitzer} photmetry with a \citet{2006Robitaille} model. The YSOs had a large spread in best-fit ages from 10$^3$-$2\times10^6$~yr. 

\begin{figure}
\includegraphics[width=\columnwidth]{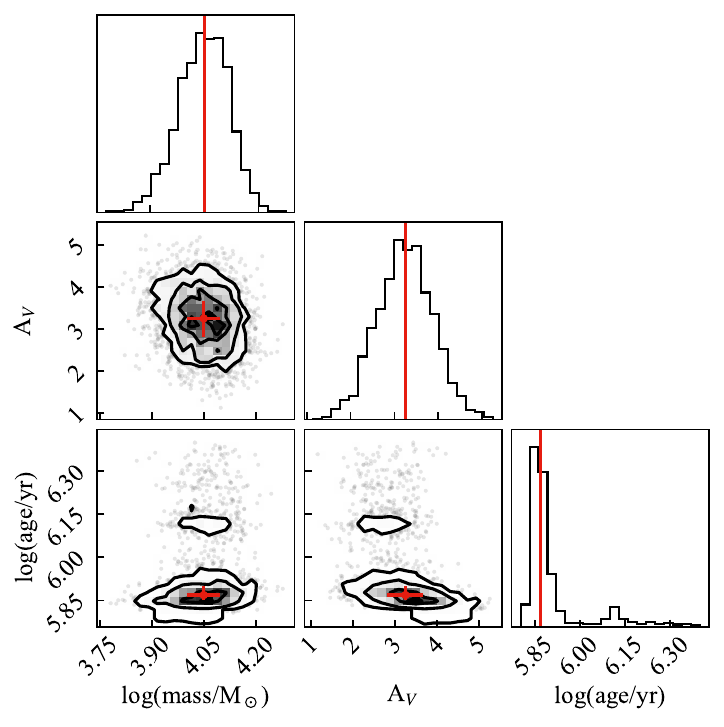}
\caption{Corner plot produced by PxP for the \citet{2009Gruendl} YSOs in N44. The red vertical lines on the histograms indicate the median of each property. The red points are the median, with the error bars representing the 25th and 75th percentiles. \label{fig:cornerN44}}
\end{figure}

\begin{figure*}
\includegraphics[width=\textwidth]{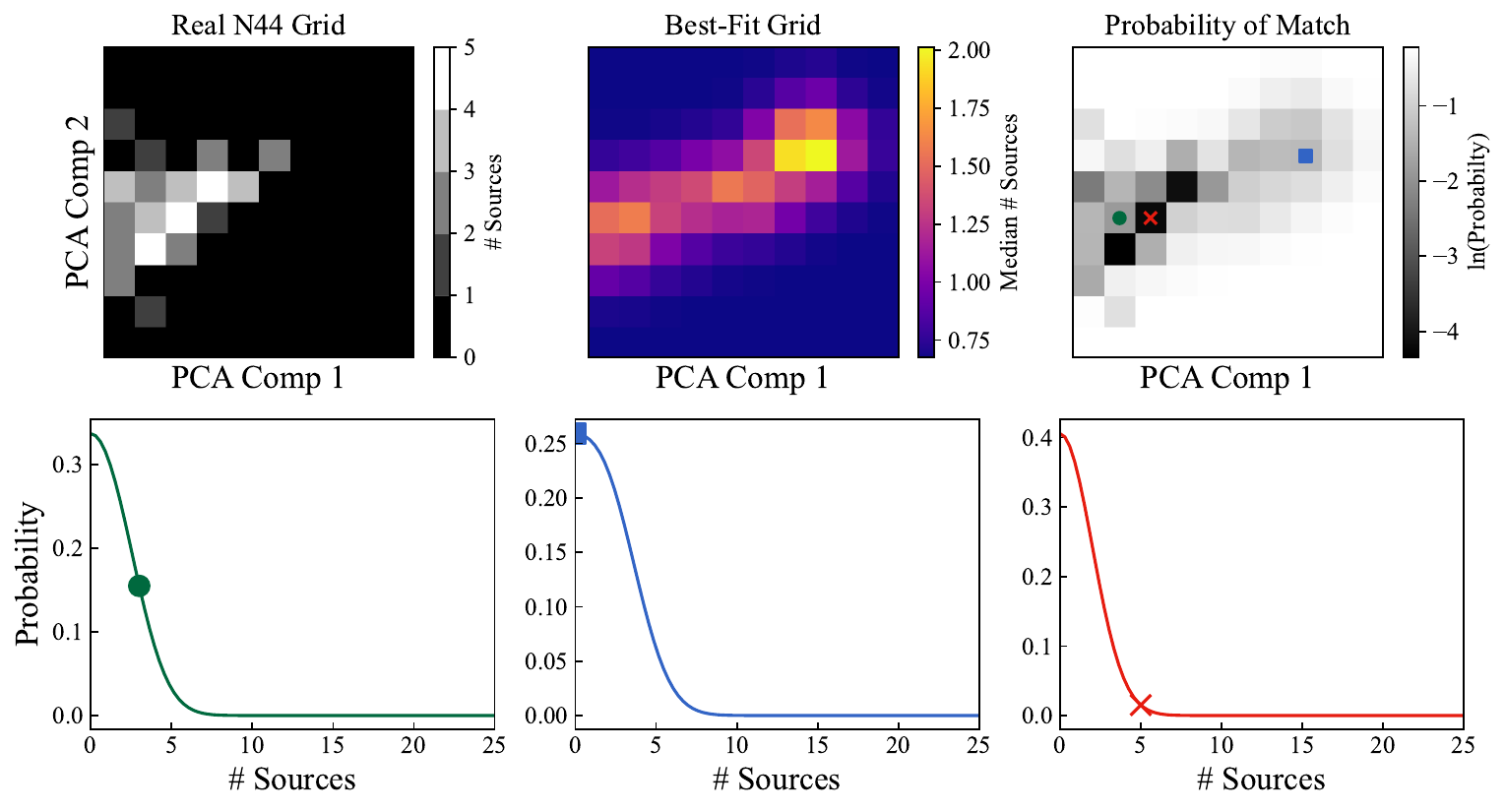}
\caption{PCA grids for the YSOs in N44, best fitting grid, and probability of match. The top left panel shows the number of YSOs with PCA components. The top middle panel shows the median number of sources in the best model with PCA components. The top right panel shows the probability the best-fit grid is a match to the real grid at each point. The bottom row shows three examples of PDFs from the best-fitting grid that are used to determine the probability grid. The grid points that the PDFs are indicated with matching color and marker in the top right panel. Based on the probabilities shown here the real YSOs in N44 are well fit by this model grid.  \label{fig:gridsN44}}
\end{figure*}

Most extragalactic star-forming regions observed by JWST will not have the level of ancillary data that is found in the LMC. Thus, for the PxP analysis, we use the older YSO catalog of \citet{2009Gruendl} identified through \textit{Spitzer} photometry, which contains flux measurements for 49 YSOs in the N44 region. After fitting these YSOs, PxP recovers a mass of (1.1$\pm0.1$)$\times10^4$~M$_\odot$, an extinction of $A_V =3.2\pm0.4$, and an age of $t_0 = 0.74^{+0.06}_{-0.03}$~Myr (Figure \ref{fig:cornerN44}). The mass is in the same order of magnitude but higher than the prediction of \citet{2012Carlson}. This higher mass likely comes from our model assumption that many of the sources are associated with unresolved clusters and binaries and not just individual YSOs, and these unresolved clusters contain more mass than a single source. The age estimate will depend heavily on the accretion history used, but PxP's estimate is in the range spanned by the \cite{2009Chen} sources. The PCA grids of the real population, the median best-fit population, and the associated probabilities are shown in Figure \ref{fig:gridsN44}, along with full PDFs for a selection of grid points. This example illustrates that PxP can effectively predict the mass and age of YSO populations from infrared SEDs. 

Using the continuous star formation history instead of the burst model leads to different estimated properties. The age estimate is 2.2$\pm0.2$~Myr, approximately $3\times$ older than the age predicted by the burst model. The mass estimate is $M_\star=(1.7\pm0.2)\times10^4$~M$_\odot$ and the derived extinction is $A_V=3.2\pm0.4$, which are comparable to the estimates from the burst model. We find that the probability of the continuous model generating the observed data is eight orders of magnitude worse than the burst model.

While the probabilities seem to show a reasonable fit to the N44 population, there is a clear distinction between the real grid and the best-fit median grid. There are a few possible reasons for this discrepancy. First, the YSO catalogs of \citet{2009Gruendl} and \citet{2012Carlson} have inconsistent SED coverage with majority of sources missing observations at short wavelengths (1.6~$\mu$m) and $\sim30$\% of sources missing observations at long wavelengths (24~$\mu$m). These inconsistent SEDs could be leading to inconsistent PCA fits shifting the number of sources contained at each grid point. Second, the completeness of the YSO catalogs could be resulting in fewer than expected faint sources being included in the population fits. Third, the \citetalias{2024Richardson} models have limited models at very high luminosities, which could lead to our model populations not accurately representing the brightest observed sources. Despite these caveats we maintain that PxP is able to predict the population parameters with relatively high accuracy when compared to previous modeling attempts.

\section{NGC~604}
We now turn our attention to NGC~604, the largest star-forming region in M33. With the new JWST images of NGC~604, we will identify YSO candidates (YSOCs) and apply PxP. Thus, obtaining a direct measurement of the ongoing star formation in NGC~604 for the first time.

\subsection{Data}
We retrieved data from the public JWST program DDT 6555 (PI:M.G.Marin). This program collected NIRCam and MIRI imaging data in a $\sim 2.5'$ region around NGC 604, a high mass star forming region in the nearby galaxy M33 \citep[$D=859~\mathrm{kpc}$;][]{2017distanceM33}. We make use of six imaging filters, three from NIRCam (F090W, F200W, and F444W) and three from MIRI (F770W, F1130W, and F1500W). The raw data were reduced using \textsc{pjpipe} developed for PHANGS-JWST \citep{2024Williams}. These reduced data are described in more detail and first appear in \citet{2025Sarbadhicary}. 

\subsection{YSOC Selection}
Before YSO Candidates (YSOCs) can be identified, we first identify all of the stellar point sources. We use the DOLPHOT software \citep{2000Dolphin,2016Dolphin} with the JWST/NIRCam module \citep{2023Weisz} in all three NIRCam bands (F090W, F200W, and F444W) simultaneously. We retain sources if they have a local signal-to-noise ratio (SNR) $>$25, (sharpness)$^2<$0.09, object type 1, and no quality flags. This catalog of NIRCam identified sources is used as a warmstart file for the DOLPHOT JWST/MIRI module \citep{2024Peltonen} to obtain photometry in the three JWST bands (F444W, F770W, F1500W). This analysis identifies 10559 point sources with fluxes in all six bands, which we assign labels of $F_x$ where $x$ is the band's central wavelength in $\mu$m. Identifying sources in NIRCam first avoids many ISM clumps being identified as point sources. This catalog may be missing some deeply embedded sources that are mid-infrared bright and near-infrared dim. However, we examined the long wavelength images from MIRI by eye and found that all discernible point sources were associated with a DOLPHOT source.
 
\begin{figure}
\includegraphics[width=\columnwidth]{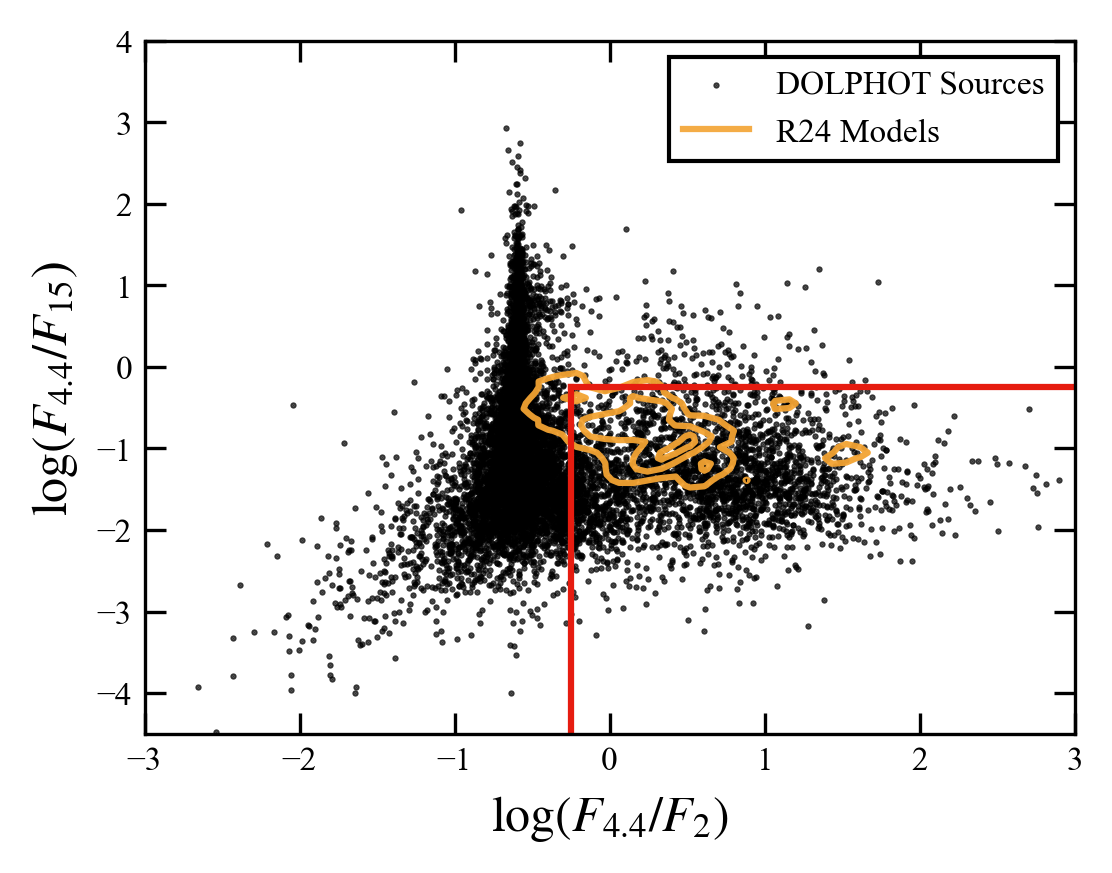}
\caption{Color-color diagram of the sources identified by DOLPHOT. The orange contours indicate the density of \citetalias{2024Richardson} models. The red lines indicate the color cut used to eliminate non-YSO contaminants. \label{fig:ccd}}
\end{figure}

The YSOC selection was performed according to the procedure outlined in \citet{2024Peltonen}. To begin eliminating contaminants we perform a cut on the color-color diagram of $\log(F_{4.4}/F_{15})$ versus $\log(F_{4.4}/F_{2})$. This color cut is shown in Figure \ref{fig:ccd} along with the location of the \citetalias{2024Richardson} models and is defined as 
\begin{equation}
    \log_{10}(F_{4.4}/F_{2})>-0.25
\end{equation}
and 
\begin{equation}
    \log_{10}(F_{4.4}/F_{15})<-0.25.
\end{equation}
The color cut on the color-color diagram leaves 2685 sources. \citet{2024Peltonen} showed on a similar color-color diagram that this removes the bulk of red supergiants and asymptotic giant branch stars, which occupy the top left and top right of the diagram, respectively. 

\begin{figure}
\includegraphics[width=\columnwidth]{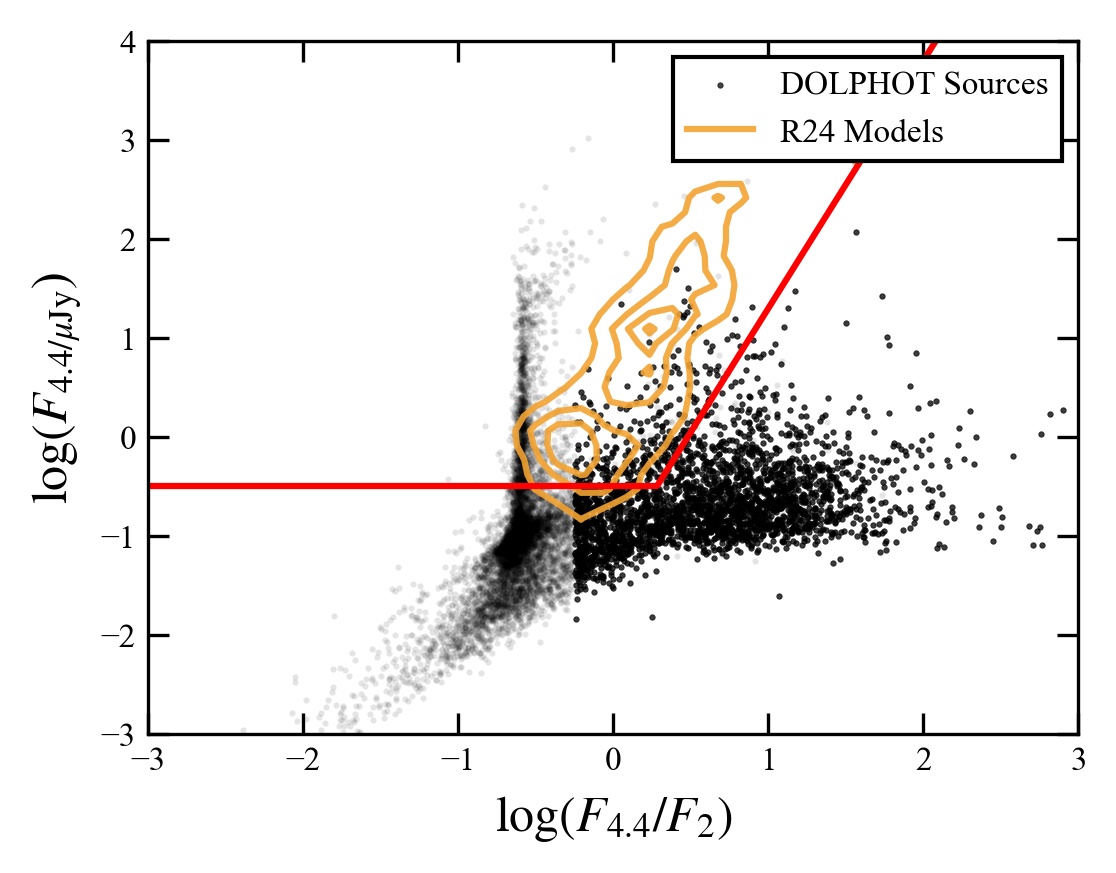}
\caption{Color-magnitude diagram of the sources identified by DOLPHOT. The sources removed by the color-color diagram cut are shown as faded symbols. The orange contours and red lines show the YSO models and color cut, respectively. \label{fig:cmd}}
\end{figure}

\begin{figure}
\includegraphics[width=\columnwidth]{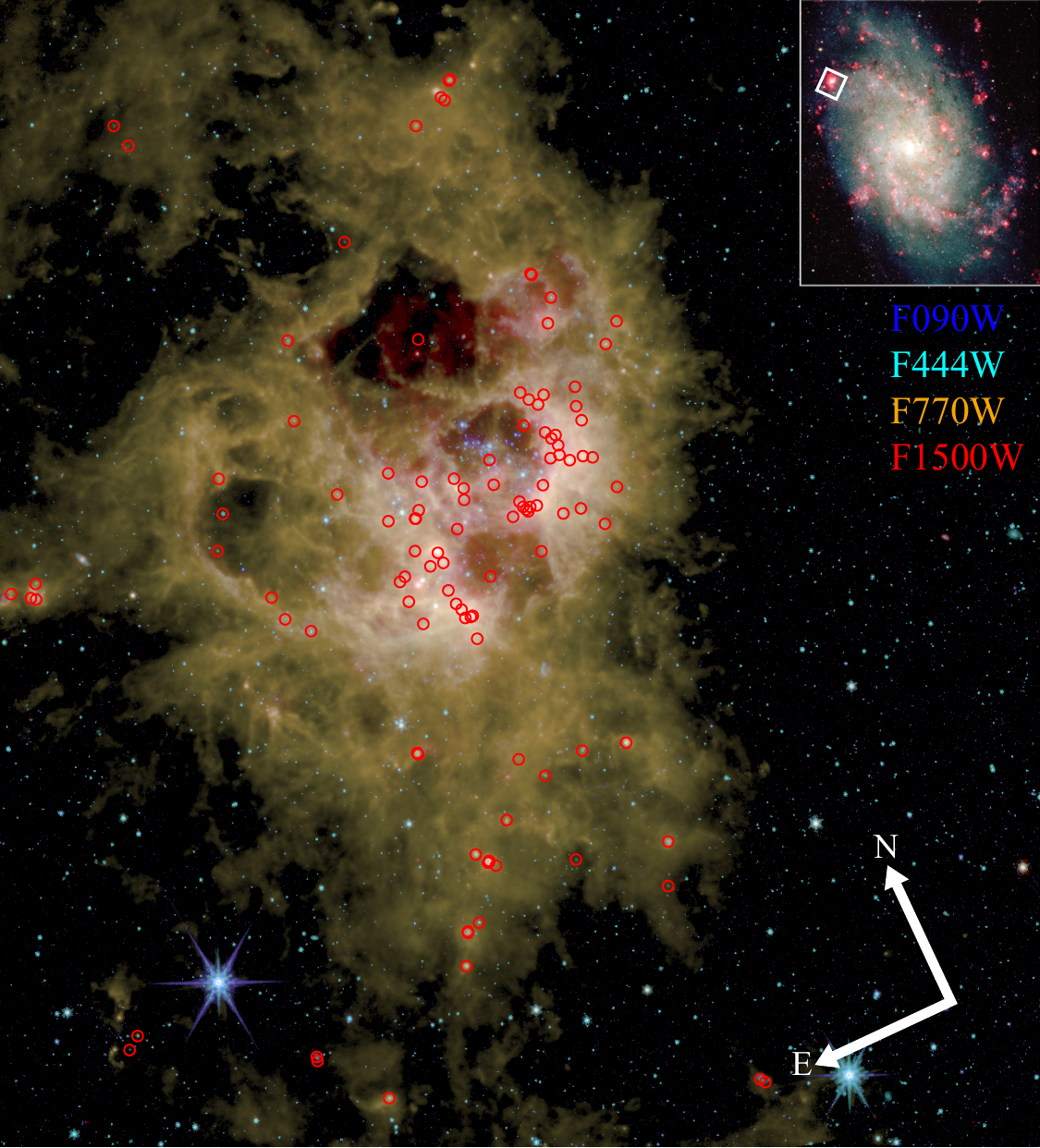}
\caption{A four color image of NGC~604 using four JWST filters (F090W, F444W, F770W, and F1500W). The red circles represent the location of the YSO candidates we identified. The top right inset shows the location of NGC~604 (white box) within a Mayall 4-meter telescope optical image of M33 \citep{2006Bband}.  \label{fig:pretty}}
\end{figure}

The next color was preformed on a color-magnitude diagram to remove additional contaminants. Figure \ref{fig:cmd} shows the $\log(F_{4.4}/\mathrm{\mu Jy})$ versus $\log(F_{4.4}/F_{2})$ along with the color cut defined as 
\begin{equation}
    \log(F_{4.4}/\mathrm{\mu Jy})>-0.5
\end{equation}
and
\begin{equation}
    \log(F_{4.4}/\mathrm{\mu Jy})>2.5 \log(F_{4.4}/F_{2}) -1.2.
\end{equation}
This color-magnitude diagram cut leaves 291 sources. An additional 26 sources were removed that had no ALMA $^{12}$CO J=1-0 detection \footnote{We use data from ALMA project 2022.1.00276.S (PI: Rosolowsky) reduced with the PHANGS-ALMA pipeline \citep{2021Leroy}.}. Finally, we perform a visual inspection on the remaining sources to remove background galaxies and ISM sources. Sources that appeared point-like in at least three filters were kept, leaving 112 YSO candidates (YSOCs). \citet{2024Peltonen} estimated the $\approx$8~percent of their YSOCs in M33 were contaminants. Therefore, we expect that $\approx$9 of our 112 YSOCs are contaminants since we followed the YSOC identification strategy of \citet{2024Peltonen} closely. In Figure \ref{fig:pretty}, we show the imaging data for the NGC 604 region with the locations of the individual YSOCs overlaid. Table \ref{tab:YSOC} lists the positions and fluxes of the YSOCs.

\begin{table*}
\centering
\caption{NGC~604 Young Stellar Object Candidates.}
 \label{tab:YSOC}
 \begin{tabular}{cccccccc}
  \hline
RA (ICRS) & DEC (ICRS) & F$_{0.9}$ ($\mu$Jy) & F$_2$ ($\mu$Jy) & F$_{4.4}$ ($\mu$Jy) & F$_{7.7}$ ($\mu$Jy) & F$_{11.3}$ ($\mu$Jy) & F$_{15}$ ($\mu$Jy) \\
\hline
23.633449 & 30.784346 & 0.05  & 0.24  & 0.54  & 0.56  & 7.82  & 226.17  \\
23.635440 & 30.782480 & 1.04  & 0.64  & 0.50 & 211.46  & 342.32  & 55.98  \\
23.635961 & 30.784565 & 0.63  & 0.81  & 0.60 & 4.25  & 2.02  & 4.81  \\
23.637317 & 30.784076 & 0.60 & 0.37  & 0.96  & 9.74  & 29.49  & 20.78  \\
23.635480 & 30.782564 & 0.62  & 0.57  & 1.60 & 29.98  & 22.27  & 117.93  \\
23.645797 & 30.783489 & 0.25  & 0.98  & 0.61  & 12.44  & 25.33  & 6.40 \\
23.653806 & 30.787200 & 0.52  & 0.45  & 0.95  & 23.47  & 36.95  & 13.15  \\
23.631949 & 30.781973 & 0.10 & 0.40 & 1.14  & 15.11  & 19.59  & 14.07  \\
23.631742 & 30.796107 & 0.01  & 0.34  & 0.72  & 15.08  & 15.57  & 13.82  \\
23.639269 & 30.782503 & 20.53  & 19.83  & 49.94  & 508.19  & 1287.14  & 1888.11  \\
  \hline
 \end{tabular}
 \\(This table is available in its entirety in a machine-readable form in the online journal.)
\end{table*}

\subsection{Model Fit}

\begin{figure}
\includegraphics[width=\columnwidth]{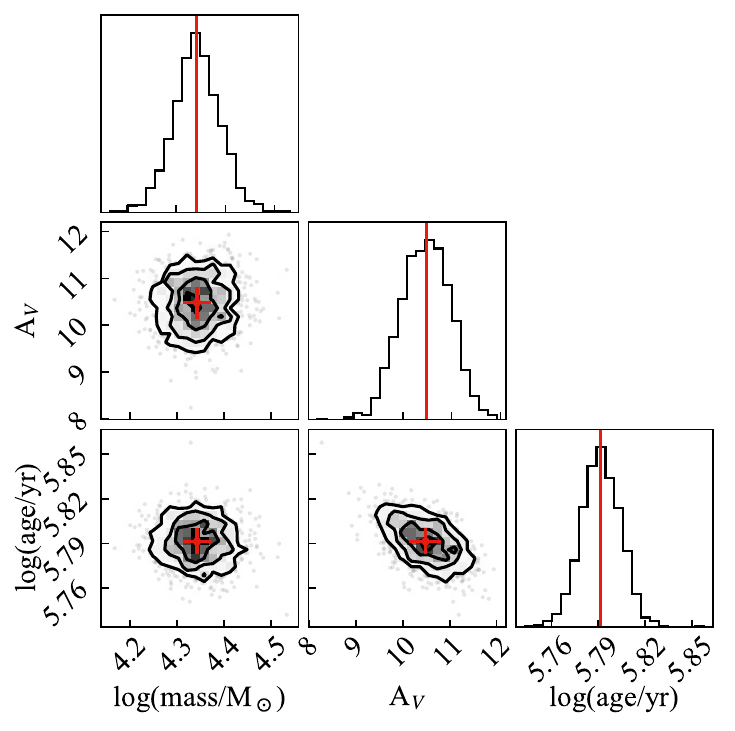}
\caption{The same as Figure \ref{fig:cornerN44} but for the YSOCs identified in NGC~604.\label{fig:corner604}}
\end{figure}

\begin{figure*}
\includegraphics[width=\textwidth]{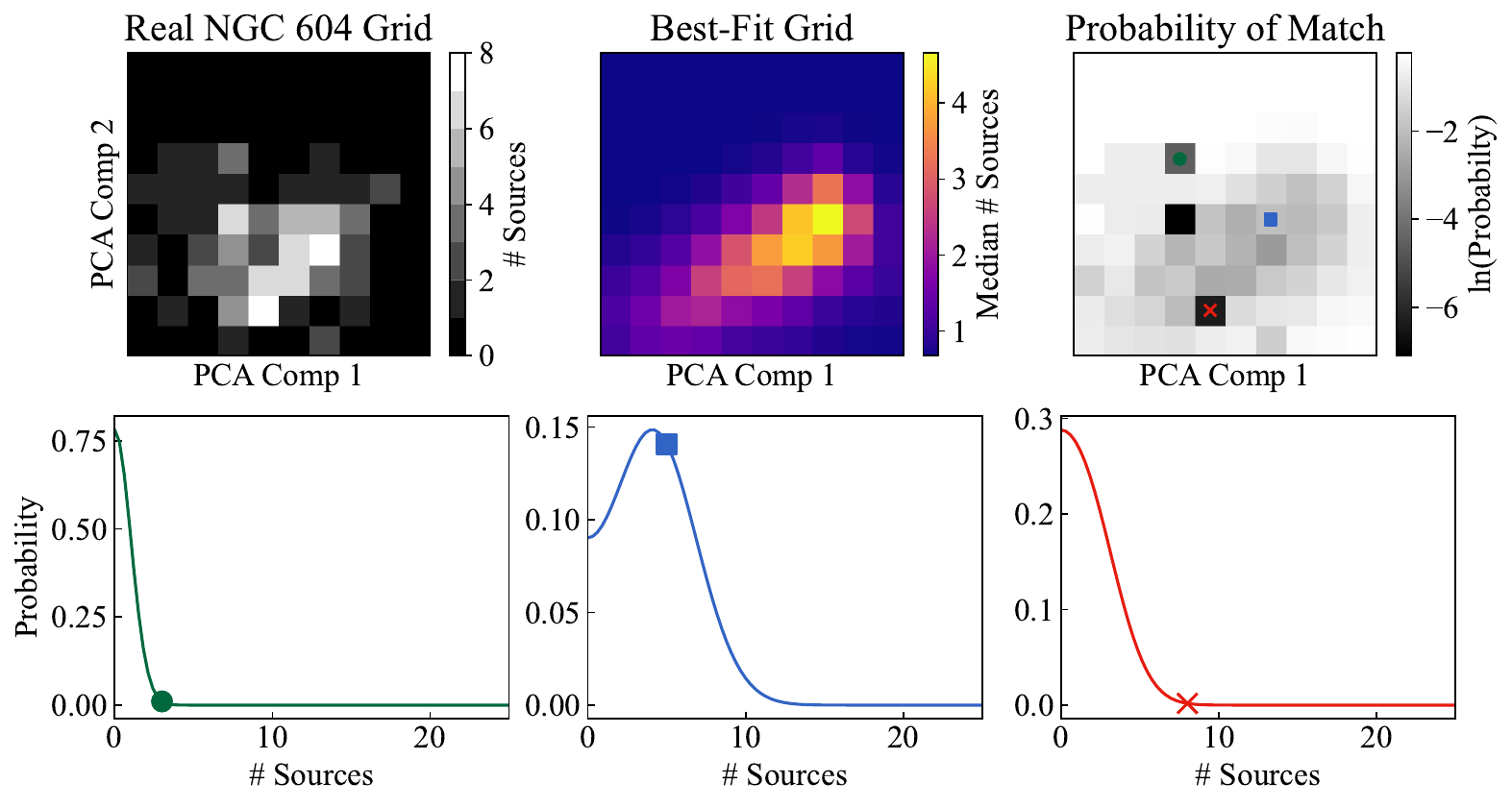}
\caption{The same as Figure \ref{fig:gridsN44} but for the YSOCs identified in NGC~604.\label{fig:grids604}}
\end{figure*}

Now equipped with a catalog of YSOCs in NGC~604, we can apply PxP. Figure \ref{fig:corner604} shows the resulting corner plot produced by PxP. From this corner plot, the best-fitting parameters for the YSOC population in NGC~604 are a mass of $M_\star = (2.2\pm0.2)\times10^4$~$M_\odot$, $A_V=10.5\pm0.3$, and $t_0 = 0.62\pm0.01$~Myr. Figure \ref{fig:grids604} shows the PCA grids from PxP for the YSOCs with the best-fitting and probability grids. The PCA grids show that PxP was able to find a reasonable fit. The grid point highlighted in red in Figure \ref{fig:grids604} shows the point with the poorest fit with more sources than would be expected by the model grid. 

\citet{2012Martinez} found a mass of $(1.2^{+1.3}_{-0.1})\times10^4$~M$_\odot$ contained in embedded star formation from model fits to unresolved \textit{Spitzer} photometry and spectroscopy.  They also found that this embedded population is a second generation of star formation, along with an existing older population ($\sim$4~Myr). The mass estimate of embedded star formation from \textit{Spitzer} is consistent with the PxP mass estimate. 

The continuous model predictions again differ from the burst model. The continuous model predicts a mass of $M_\star = (3.6\pm0.3)\times10^4$~$M_\odot$, $A_V=8.8\pm0.4$, and $t_0 = 1.9\pm0.4$~Myr. Again the continuous model is a worse fit with the total probability being 10 orders of magnitude worse than the burst model.

Now with an estimated age and mass of the ongoing star formation in NGC~604, we can calculate the local star formation rate. Simply dividing the total mass of the population by the age of the population gives a star formation rate of 0.035$\pm0.003$~$M_\odot$/yr. The star formation maps of \citet{2015Boquien} predict the star formation rate across M33 in a variety of tracers. The predicted star formation rate in NGC~604 using H$\alpha$ is 0.016$\pm0.002$~$M_\odot$/yr \citep{2015Boquien}. H$\alpha$ should be tracing a later stage of star formation likely corresponding to the first generation of star formation seen by \citet{2012Martinez}. If we consider the continuous model the greater age gives a star formation rate of 0.019$\pm0.002$~$M_\odot$/yr closer but still larger than the \citep{2015Boquien} estimate. Therefore, our estimate implies that star formation activity has not decreased with the second generation and has possibly even increased.

For both the N44 and the NGC 604 data, the continuous model finds that the ages are older by a factor of 3 and masses larger by a factor of 1.5 when compared to the burst model. From our definition of a continuous star formation history with oldest star having age of $t_0$, the mean age of a star will be $t_0/2$. Naively, a burst model should best match this mean age, so we would expect the burst ages to be a factor of two shorter than the continuous age. That we see a factor of 3 shorter in the two populations we study here suggests the true star formation history may be accelerating in these system \citep{2000palla}. The difference in mass estimate likely arises because the average $L_\mathrm{IR}/M$ ratio of a continuously-forming population decreases with population age in our models. While no conclusions can be rigorously drawn from a sample of two regions, this analysis does illustrate the potential for using PxP to assess the changing rates of star formation across a wide range of systems.

\section{Conclusion}
We have developed a comprehensive framework that can predict the age and mass of an observed population of YSOs. Utilizing the latest radiative transfer models of \citetalias{2024Richardson} and accretion histories of \citet{2011Offner}, we create model populations of YSOs. This population synthesis code includes binaries and unresolved clusters with simulated intracluster medium. We then fit these simulated populations to real populations with PCA and maximum likelihood fitting in a complete framework we have dubbed PxP.

PxP was tested through self-fitting and a known population of YSOs. First, self-fitting tests show that PxP is most effective at predicting the mass of YSO populations. While the age estimations are much more uncertain, they become more accurate with a greater number of sources ($\gtrsim40$). Applying PxP to the well-studied N44 region gives a mass of (1.1$\pm0.1$)$\times10^4$~M$_\odot$ and an age of $0.74^{+0.06}_{-0.03}$~Myr. These results are consistent with previous studies of this region.

Finally, we identify a new population of YSO candidates in NGC~604 that can then have its age and mass predicted with PxP. We identified all of the point sources in the archival JWST data in NGC~604. 112 YSO candidates were found from these point sources through color cuts and visual inspection. PxP was applied to this population to retrieve a mass of $(2.2\pm0.2)\times10^4$~$M_\odot$ and an age of $0.62\pm0.01$~Myr. These findings are consistent with estimations from low-resolution observations. 

In this paper we have demonstrated the basic utility of PxP utilizing a specific set of models. This framework can be relaxed and expanded, notably using different star formation histories and accretion histories. With sufficient numbers of sources and regions, we even have the opportunity to evaluate what star formation histories or accretion histories are most consistent with the observed data. This method will allow for the star formation rates to be predicted in a broader range of environments with greater accuracy than ever before. 

Future work will apply PxP to the well-studied clouds in the solar neighborhood. This analysis will ensure that PxP can estimate the properties of populations only containing low-mass YSOs. A full public release of PxP is planned to accompany this Milky Way analysis. 

\software{emcee \citep{2013emcee}; astropy \citep{astropy}; JWST pipeline \citep{jwstpipeline}; pjpipe \citep{2024Williams}; corner \citep{corner}; matplotlib \citep{matplotlib}; dolphot \citep{2016Dolphin}}

\facilities{Spitzer, JWST, ALMA}

\section*{Acknowledgments}
We are grateful for the anonymous reviewer who provided insightful comments that improved the quality of this manuscript. All of the JWST data presented in this paper were obtained from the Mikulski Archive for Space Telescopes (MAST) at the Space Telescope Science Institute. The specific observations analyzed can be accessed via \dataset[https://doi.org/10.17909/h61s-7n17]{https://doi.org/10.17909/h61s-7n17}. STScI is operated by the Association of Universities for Research in Astronomy, Inc., under NASA contract NAS5–26555. Support to MAST for these data is provided by the NASA Office of Space Science via grant NAG5–7584 and by other grants and contracts. 

This paper makes use of the following ALMA data: ADS/JAO.ALMA\#2022.1.00276.S. ALMA
is a partnership of ESO (representing its member states), NSF (USA) and NINS (Japan),
together with NRC (Canada), MOST and ASIAA (Taiwan), and KASI (Republic of Korea), in
cooperation with the Republic of Chile. The Joint ALMA Observatory is operated by
ESO, AUI/NRAO and NAOJ. The National Radio Astronomy Observatory is a facility of the National Science Foundation operated under cooperative agreement by Associated Universities, Inc.

JP and ER acknowledge support from the Natural Science and Engineering Research Council Canada, Funding Reference RGPIN-2022-03499 and from the Canadian Space Agency, Funding Reference numbers 22JWGO1-20, 23JWGO2-A08. AG acknowledges support from the NSF under grants AST 2008101 and CAREER 2142300.

\bibliography{PASPsample631}{}
\bibliographystyle{aasjournal}



\end{document}